\RequirePackage{etoolbox}
\csdef{input@path}{{style/}{graphics/}}
\documentclass[aos,MSNbibl,nameyear,seceqn,dvips]{arximspdf}
\usepackage[mathscr]{euscript}
\usepackage{graphicx}

\doi{10.1214/14-AOS1297}
\volume{43}
\issue{2}
\pubyear{2015}
\firstpage{713}
\lastpage{740}
\docsubty{FLA}

\makeatletter
\newcommand{\ds}{\displaystyle}
\newcommand{\eqref}[1]{(\ref{#1})}
\newtheorem{theorem}{Theorem}[section]

\newtheorem{corollary}[theorem]{Corollary}

\newproclaim{definition}{Definition}[section]
\newproclaim{remark}{Remark}[section]
\newproclaim{example}{Example}
\newproclaim{model}{Model}
\newproclaim{claim}{Claim}
\makeatother

\begin{document}
\begin{frontmatter}

\title{Detecting gradual changes in locally stationary~processes\thanksref{T1}}
\runtitle{Detecting gradual changes}

\begin{aug}
\author[A]{\fnms{Michael}~\snm{Vogt}\ead[label=e1]{michael.vogt@uni-konstanz.de}}
\and
\author[B]{\fnms{Holger}~\snm{Dette}\corref{}\ead[label=e2]{holger.dette@ruhr-uni-bochum.de}}
\runauthor{M. Vogt  and H. Dette}
\affiliation{Universit\"{a}t Konstanz and Ruhr-Universit\"{a}t Bochum}
\address[A]{Department of Mathematics and Statistics \\
Universit\"at Konstanz\\
78457 Konstanz\\
Germany \\
\printead{e1}}
\address[B]{Department of Mathematics\\
Ruhr-Universit\"{a}t Bochum\\
44780 Bochum\\
Germany\\
\printead{e2}}
\end{aug}
\thankstext{T1}{Supported in part by the Collaborative Research
Center ``Statistical modeling of nonlinear dynamic processes'' (SFB
823, Teilprojekt A1, C1) of the German Research Foundation.}

%
\received{\smonth{3} \syear{2014}}
%
\revised{\smonth{12} \syear{2014}}

\begin{abstract}
In a wide range of applications, the stochastic properties of the
observed time series change over time. The changes often occur
gradually rather than abruptly: the properties are (approximately)
constant for some time and then slowly start to change. In many cases,
it is of interest to locate the time point where the properties start
to vary. In contrast to the analysis of abrupt changes, methods for
detecting smooth or gradual change points are less developed and often
require strong parametric assumptions. In this paper, we develop a
fully nonparametric method to estimate a smooth change point in a
locally stationary framework. We set up a general procedure which
allows us to deal with a wide variety of stochastic properties
including the mean, (auto)covariances and higher moments. The
theoretical part of the paper establishes the convergence rate of the
new estimator. In addition, we examine its finite sample performance by
means of a simulation study and illustrate the methodology by two
applications to financial return data.
\end{abstract}

\end{frontmatter}

\section{Introduction}

In many applications, the stochastic properties of the observed time
series such as the mean, the variance or the distribution change over
time. In the classical structural break setting, the changes are
abrupt: the stochastic properties are constant for some time and then
suddenly jump to another value. In a number of situations, however, the
changes occur gradually rather than abruptly: the properties are
(approximately) constant for a while and then reach a time point where
they slowly start to change. We refer to this time point as a smooth or
gradual change point in what follows.

\begin{figure}

\includegraphics{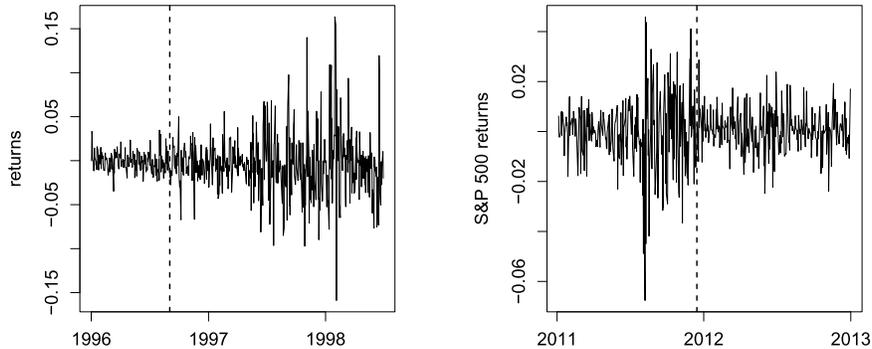}

\caption{The left-hand panel shows the daily returns of the Thai stock
index from January 1996 to June 1998. The right-hand panel depicts the
daily returns of the S\&P 500 index from the beginning of 2011 to the
end of 2012. The vertical dashed lines indicate the gradual change points
estimated by the method developed in this paper.}\label{fig1}
\end{figure}

Locating a smooth change point is important in a wide range of
applications. A~first example concerns the Asian financial crisis in
1997. Economists are strongly interested in better understanding the
dynamics of this crisis. To do so, it is crucial to locate the time
points when the various countries of the East Asian region became
affected by the crisis. This issue can be tackled by analyzing the
volatility levels of the East Asian stock indices: an increase in
volatility of a country's index indicates that this country gets
affected. In many cases, the increase occurs gradually rather than
abruptly: the volatility level slowly rises as the country's economic
situation deteriorates. This is illustrated by the left-hand panel of
Figure~\ref{fig1}, which shows the daily returns of the Thai stock
index. As can be seen, the volatility level is fairly constant at the
beginning of the sample and then starts to gradually build up. We thus
face the problem of locating a smooth change point in the volatility level.

A second example concerns forecasting. The right-hand panel of Figure~\ref{fig1} shows the daily returns of the S\&P 500 stock index from the
beginning of 2011 to the end of 2012. Inspecting the data, it is
apparent that the volatility level changes over time. Moreover, the
plot suggests that the volatility is roughly constant in 2012 but
gradually increases before that. Denoting the present time point by
$T$, practitioners are often interested in identifying the time
interval $[t_0,T]$ where the volatility level is more or less constant.
Put differently, they are interested in localizing the time point $t_0$
prior to which the volatility starts to substantially vary over time.
Once the time point $t_0$ has been identified, it is common practice in
volatility forecasting to fit a model to the data in the time span
$[t_0,T]$; see, for example, \citet{Chen2010}.

Other examples can be found in a variety of different fields. In
climatology, it is of interest to locate changes in the mean, variance
or covariance structure of certain climate processes. These processes
often change their behavior gradually rather than abruptly. The issue
is thus to locate smooth change points in them. \citet{Mallik2011}, for
instance, consider the problem of detecting a smooth change point in
global temperature anomaly records. A final example comes from
neuroscience. In the analysis of EEG data, a question of interest is to
locate the time point where an epileptic seizure occurs. The onset of a
seizure arguably coincides with a change in the autocovariance
structure of the EEG data. The aim is thus to estimate the time point
where the autocovariance structure starts to vary.

In most applications, there is not much known about the way in which
the stochastic properties of interest evolve over time, once the smooth
change point has been reached. For instance, there is no economic
theory suggesting that the increase of the volatility level in the
return series of Figure~\ref{fig1} should have a specific parametric
form. It is thus important to have flexible nonparametric methods at
hand which allow us to locate a smooth change point without imposing
strong parametric restrictions on the time-varying behavior of the
stochastic properties under consideration.

The main goal of this paper is to develop such a method. More
precisely, we tackle the following estimation problem: suppose we
observe a sample of data $\{ X_{t,T} \dvtx  t=1,\ldots,T \}$ and are
interested in a stochastic feature such as the mean $\mathbb
{E}[X_{t,T}]$ or
the variance $\operatorname{Var}(X_{t,T})$. Moreover, assume that the
feature is
time-invariant on the time span $\{ 1, \ldots, t_0 \}$, or
equivalently, on the rescaled time interval $[0,u_0]$ with $u_0 =
t_0/T$ and then starts to gradually vary over time. Our aim is to
estimate the rescaled time point $u_0$. We do not impose any parametric
restrictions on the time-varying behavior of the feature of interest
after $u_0$. In this sense, our model setting is completely
nonparametric. Moreover, rather than restricting attention to a
specific stochastic property, we set up a general procedure which
allows us to deal with a wide variety of features including the mean,
(auto)covariances and higher moments of the time series at hand. We
tackle the problem of estimating $u_0$ within a locally stationary
framework which is well suited to model gradual changes and is formally
introduced in Section~\ref{sec-loc-stat}.

The nonparametric nature of our estimation problem sharply
distinguishes it from standard change point problems and requires new
methodology. The literature commonly imposes strong parametric
restrictions on the time-varying behavior of the stochastic properties
at hand. In the vast majority of papers, the changes are abrupt, that
is, the properties are assumed to be constant over time apart from some
occasional structural breaks. The detection of sudden structural breaks
has a long history originating from quality control; see, for example,
\citeauthor{page1954} (\citeyear{page1954,page1955}) for some early references. Since then a vast
literature on break detection has evolved; see \citet{horkokste1999},
\citet{aueetal2009} or \citet{davis2006}, among many others.

The literature on detecting gradual changes is much more scarce than
that on abrupt changes. Most references consider location models of a
very simple parametric form. Several authors investigate broken line
regression models with independent normally distributed errors [see,
e.g., \citet{hinkley1970} or \citet{siezha1994}] and the performance of
control charts under a gradual change in the mean [see \citeauthor{bissel1984a} (\citeyear{bissel1984a,bissel1984b}) or
\citeauthor{gan1991} (\citeyear{gan1991,gan1992})]. Other work
considers estimators and tests in models where the linear drift has
been replaced by some smooth parametric function (such as a
polynomial) and the errors are assumed to be i.i.d. but not
necessarily normal; see \citet{huskova1999}, \citet{husste2002} and also
\citet{aueste2002} for a generalization to the dependent case.

More recently, there has been some work on the problem of detecting
smooth change points in some simple nonparametric settings. Most
authors consider the location model $X_{t,T} = \mu(\frac{t}{T}) +
\varepsilon_t$ with zero mean i.i.d. errors $\varepsilon_t$. Indeed,
in many cases, the errors are even assumed to be Gaussian. Suppose that
the mean function $\mu$ is constant on the interval $[0,u_0]$, that is,
$\mu(u) = \overline{\mu}$ for $u \le u_0$, and then starts to smoothly
vary over time. Under appropriate smoothness conditions, $u_0$ can be
regarded as a break point in the $k$th derivative of $\mu$. It can thus
be estimated by methods to detect a break point in a higher-order
derivative of a nonparametric function; see \citet{Mueller1992} for an
early reference and, for example, \citet{Raimondo1998} and \citet
{Goldenshluger2006} who derive minimax rates in the model with Gaussian
errors. \citet{Mallik2011} and \citet{Mallik2013} propose an alternative
$p$-value based approach to estimate $u_0$ when $\mu$ is a smooth
nonparametric function that is restricted to take values larger than
$\overline{\mu}$ at time points $u > u_0$, that is, $\mu(u) >
\overline
{\mu}$ for $u > u_0$. Finally, \citet{Mercurio2004} study sequential
testing procedures for change point detection in some simple
nonparametric volatility models. All these methods are tailored to a
very specific model setting and often rely on strong distributional
assumptions. Our procedure in contrast is very general in nature and
can be applied to a wide variety of settings. Moreover, it does not
rely on any distributional restrictions. In the location model $X_{t,T}
= \mu(\frac{t}{T}) + \varepsilon_t$, for instance, we do not even
require the errors to be independent or stationary. In fact, we are
able to estimate $u_0$ as long as the errors are locally stationary.

In Section~\ref{sec-est}, we introduce our estimator of the time point
$u_0$, which is based on a refinement of the CUSUM principle. To
construct it, we proceed in two steps. In the first, we set up a
function $\mathscr{D}\dvtx  [0,1] \rightarrow\mathbb{R}_{\ge0}$, where
$\mathscr{D}(u)$
measures the amount of time-variation in the stochastic feature of
interest within the interval $[0,u]$. By construction, $\mathscr{D}(u)
= 0$
if there is no time-variation on the interval $[0,u]$ and $\mathscr
{D}(u) >
0$ if there is some time-variation involved. Since $\mathscr{D}$ is not
observed, we replace it by an estimator $\hat{\mathscr{D}}_T$.
Section~\ref{sec-measure} gives a detailed account of how to construct the measure
of time-variation $\mathscr{D}$ and its estimator $\hat{\mathscr
{D}}_T$. The time
point $u_0$ can now be characterized as the point where the measure
$\mathscr{D}$ starts to deviate from zero. This characterization is
used in
the second step to come up with an estimator of $u_0$. Section~\ref{sec-est} describes in detail how to set up this estimator.

In Section~\ref{sec-asym}, we derive the convergence rate of the
proposed estimator. As we will see, the rate depends on the degree of
smoothness of the stochastic feature of interest at $u_0$. This
reflects the fact that it becomes harder to locate the time point $u_0$
when the feature varies more slowly and smoothly around this point. Our
method depends on a tuning parameter which is similar in nature to the
critical value of a test procedure and which can be chosen to keep a
pre-specified probability of underestimating the point $u_0$. We derive
a data driven choice of the tuning parameter with good theoretical and
practical properties in Sections~\ref{subsec-asym-threshold} and \ref
{sec-impl}. The first and second part of Section~\ref{sec-sim}
investigate the small sample performance of our method by means of a
simulation study and compare it with competing methods for the location
model $X_{t,T} = \mu(\frac{t}{T}) + \varepsilon_t$. Additional
simulations can be found in the supplement [\citet{vogtdett2014}]. In the
third part of Section~\ref{sec-sim}, we use our method to analyze some
data from the 1997 Asian crisis and the S\&P 500 returns from Figure~\ref{fig1}. Finally, in Section~\ref{sec-ext}, we show that our method
is not only able to detect smooth change points but also abrupt breaks.
Hence, it is not only useful in applications where the change point to
be detected is gradual. It also provides a robust way to estimate a
change point when it is not clear whether the change occurs gradually
or abruptly.

\section{Model setting}\label{sec-loc-stat}
\setcounter{equation}{0}

Throughout the paper, we assume that the sample of observations $\{
X_{t,T}\dvtx  t=1,\ldots,T \}$ comes from a locally stationary process of
$d$-dimensional variables $X_{t,T}$. Specifically, we work with the
following concept of local stationarity, which was introduced in \citet
{Vogt2012}.

\begin{definition}\label{def-loc-stat}
The array $\{ X_{t,T}\dvtx  t=1,\ldots,T \}_{T=1}^{\infty}$ is called a
locally stationary process if for each rescaled time point $u \in
[0,1]$, there exists a strictly stationary process $\{ X_t(u)\dvtx  t \in
\mathbb{Z}\}$ with the property that
\[
\bigl\| X_{t,T} - X_t(u) \bigr\| \leq \biggl( \biggl| \frac{t}{T}-u
\biggr| + \frac{1}{T} \biggr) U_{t,T}(u) \qquad\mbox{a.s.}
\]
Here, $\| \cdot\|$ denotes a norm on $\mathbb{R}^d$, and $\{ U_{t,T}(u)\dvtx
t=1,\ldots,T \}_{T=1}^{\infty}$ is an array of positive random
variables whose $\rho$th moment is uniformly bounded for some $\rho>
0$; that is, $\mathbb{E}[U^\rho_{t,T}(u)] \le C < \infty$ for some fixed
constant $C$.
\end{definition}

Our definition of local stationarity is similar to those in
\citet{Dahlhaus2006} and \citet{Koo2012}, for example. The intuitive idea
behind it is that a process is locally stationary if it behaves
approximately stationary locally in time, that is, over short time
periods. This idea is turned into a rigorous concept by requiring that
locally around each rescaled time point $u$, the process $\{ X_{t,T} \}
$ can be approximated by a stationary process $\{ X_t(u) \}$ in a
stochastic sense.

The concept of local stationarity relies on rescaling time to the unit
interval. The main reason for doing so is to obtain a reasonable
asymptotic theory. Rescaling the time argument is also common in the
investigation of change points. While a completely specified parametric
model as considered in \citet{hinkley1970} or \citet{siezha1994} does not
need this technique, more general approaches are usually based on
rescaling arguments; see \citet{huskova1999} or \citet{aueste2002},
among others.

Let $\lambda_{t,T}$ be some time-varying feature of the locally
stationary process $\{ X_{t,T} \}$ such as the mean $\mathbb
{E}[X_{t,T}]$ or
the variance $\operatorname{Var}(X_{t,T})$. Generally speaking, we
allow for any
feature $\lambda_{t,T}$ which fulfills the following property:
\begin{longlist}[($P_{\lambda}$)]
\item[($P_{\lambda}$)] $\lambda_{t,T}$ is uniquely determined by the
set of moments $\{ \mathbb{E}[f(X_{t,T})]\dvtx  f \in\mathcal{F} \}$ in
the sense
that there is a one-to-one correspondence between $\lambda_{t,T}$ and
this set, where $\mathcal{F}$ is a family of measurable functions $f\dvtx
\mathbb{R}^d \rightarrow\mathbb{R}$.
\end{longlist}
Note that ($P_{\lambda}$) is a fairly weak condition which is satisfied
by a wide range of stochastic features. Indeed, it essentially allows
us to deal with any feature that can be expressed in terms of a set of
moments. We illustrate the property ($P_{\lambda}$) with some examples:

\renewcommand{\theexample}{I}
\begin{example} \label{example1}
Let $\lambda_{t,T}$ be the mean $\mu_{t,T} = \mathbb{E}[X_{t,T}]$ of a
univariate locally stationary process $\{X_{t,T}\}$. Then the
corresponding family of functions is simply $\mathcal{F} = \{ \mbox{id}
\}$, since the mean $\mu_{t,T}$ can be written as $\mathbb{E}[\mbox
{id}(X_{t,T})]$.
\end{example}

\renewcommand{\theexample}{II}
\begin{example}\label{example2}
Let $\lambda_{t,T}$ be the vector of the first $p$ autocovariances of a
univariate locally stationary process $\{ Y_{t,T} \}$ whose elements
$Y_{t,T}$ are centered for simplicity. Specifically, define $\gamma
_{\ell,t,T} = \operatorname{Cov}(Y_{t,T},Y_{t-\ell,T})$ to be the
$\ell$th order
autocovariance, and set $\lambda_{t,T} = (\gamma_{0,t,T},\ldots
,\gamma
_{p,t,T})^{\intercal}$. To handle this case, we regard the data as
coming from the $(p+1)$-dimensional process $\{ X_{t,T} \}$ with
$X_{t,T} = (Y_{t,T},Y_{t-1,T},\ldots,Y_{t-p,T})^{\intercal}$. We now
define functions $f_{\ell}\dvtx  \mathbb{R}^{p+1} \rightarrow\mathbb{R}$
for $0 \le
\ell\le p$ by $f_{\ell}(x) = x_0 x_{\ell}$, where $x = (x_0,\ldots
,x_p)^{\intercal}$. As $\mathbb{E}[f_{\ell}(X_{t,T})] = \mathbb{E}[Y_{t,T}
Y_{t-\ell,T}] = \gamma_{\ell,t,T}$, we obtain that $\mathcal{F} = \{
f_0,\ldots,f_p \}$ in this setting.
\end{example}

\renewcommand{\theexample}{III}
\begin{example}\label{example3}
Consider a $d$-dimensional locally stationary process $\{X_{t,T}\}$
whose elements $X_{t,T} = (X_{t,T,1},\ldots,X_{t,T,d})^{\intercal}$
are again centered for simplicity. Let $\lambda_{t,T}$ be the vector of
covariances $\nu_{t,T}^{(i,j)} = \operatorname
{Cov}(X_{t,T,i},X_{t,T,j})$, that is,
$\lambda_{t,T} = (\nu_{t,T}^{(i,j)})_{1 \le i \le j \le d}$.
Analogously\vspace*{1pt} as in the previous example, $\mathcal{F} = \{ f_{ij} \dvtx  1
\le i \le j \le d \}$ with $f_{ij}(x) = x_i x_j$.
\end{example}

We next define $\lambda(u)$ to be the stochastic feature of the
approximating process $\{X_t(u)\}$ which corresponds to $\lambda
_{t,T}$. This means that $\lambda(u)$ is fully characterized by the set
of moments $\{ \mathbb{E}[f(X_t(u))]\dvtx  f \in\mathcal{F} \}$.
Throughout the
paper, we assume that
%
\begin{equation}
\label{gen-measure-1}
\sup_{f \in\mathcal{F}} \bigl| \mathbb{E}\bigl[f(X_{t,T})
\bigr] - \mathbb {E}\bigl[f\bigl(X_t(u)\bigr)\bigr] \bigr| \le C \biggl( \biggl|
\frac{t}{T} - u \biggr| + \frac{1}{T} \biggr),
\end{equation}
which is implied by the high-order condition (C4) in Section~\ref{subsec-asym-ass}. In a wide range of cases, inequality \eqref
{gen-measure-1} boils down to mild moment conditions on the random
variables $X_{t,T}$, $X_t(u)$ and $U_{t,T}(u)$. This in particular
holds true in Examples \ref{example1}--\ref{example3} as discussed in
Section~\ref{subsec-asym-ass}. Inequality \eqref{gen-measure-1}
essentially says that $\lambda_{t,T}$ and $\lambda(u)$ are close to
each other locally in time. In the time-varying mean setting from
Example~\ref{example1}, it can be expressed as $| \mu_{t,T} - \mu(u)|
\le C (|\frac{t}{T} - u| + \frac{1}{T})$ with $\mu(u)$ being the mean
of $X_t(u)$. In Example~\ref{example2}, it is equivalent to the
statement $\| (\gamma_{0,t,T},\ldots,\gamma_{p,t,T})^{\intercal} -
(\gamma_{0}(u),\ldots,\gamma_{p}(u))^{\intercal} \| \le C (|\frac
{t}{T} - u| + \frac{1}{T})$, where $\gamma_{\ell}(u) = \operatorname{Cov}
(Y_t(u),Y_{t-\ell}(u))$ and $\| \cdot\|$ is some norm on $\mathbb{R}
^{p+1}$. Similarly, in Example~\ref{example3}, it says that $\| (\nu
_{t,T}^{(i,j)})_{i,j=1,\ldots,d} - (\nu^{(i,j)}(u))_{i,j=1,\ldots,d}
\|
\le C (|\frac{t}{T} - u| + \frac{1}{T})$, where $\nu^{(i,j)}(u) =
\operatorname{Cov}
(X_{t,i}(u),X_{t,j}(u))$. Hence, if \eqref{gen-measure-1} holds true,
the feature $\lambda_{t,T}$ converges to $\lambda(u)$ locally in time.
In particular, time-variation in $\lambda_{t,T}$ is asymptotically
equivalent to time-variation in $\lambda(u)$. To detect whether the
stochastic feature $\lambda_{t,T}$ of interest changes over time, we
may thus check for variations in the approximating quantity $\lambda(u)$.

Our estimation problem can now be formulated as follows: assume that
$\lambda(u)$ does not vary on the rescaled time interval $[0,u_0]$ but
is time-varying after $u_0$. Our aim is to estimate the time point
$u_0$ where $\lambda(u)$ starts to change over time.

\section{A measure of time-variation}\label{sec-measure}
\setcounter{equation}{0}

In this section, we set up a function $\mathscr{D}\dvtx  [0,1] \rightarrow
\mathbb{R}
_{\ge0}$ which captures time-variation in the stochastic feature
$\lambda= \lambda(\cdot)$ of interest and explain how to estimate it.
By construction, the function $\mathscr{D}$ has the property
\renewcommand{\theequation}{$P_{\mathscr{D}}$}
%
\begin{equation}\label{PD}
\mathscr{D}(u) \cases{= 0, & $\quad\mbox{if } \lambda\mbox{ does not vary on }
[0,u]$, \vspace*{3pt}
\cr
> 0, & $\quad\mbox{if } \lambda\mbox{ varies on }
[0,u]$,}
\end{equation}
and is called a measure of time-variation. In what follows, we describe
how to set up such a measure for a generic stochastic feature that
satisfies ($P_{\lambda}$).

Our construction is based on the following idea: by the property
($P_{\lambda}$), the feature $\lambda(w)$ is fully characterized by the
values $\mathbb{E}[f(X_t(w))]$ with $f$ running over all functions in the
family $\mathcal{F}$. This implies that time-variation in $\lambda(w)$
is equivalent to time-variation in the moments $\mathbb{E}[f(X_t(w))]$ for
some $f \in\mathcal{F}$. To detect changes in $\lambda(w)$ over time,
we may thus set up a function which captures time-variation in the
quantities $\mathbb{E}[f(X_t(w))]$ for any $f \in\mathcal{F}$. This idea
underlies the following definition:
\renewcommand{\theequation}{\thesection.\arabic{equation}}
\setcounter{equation}{0}
\begin{equation}
\label{Dsup-gen}
\mathscr{D}(u) = \sup_{f \in\mathcal{F}} \sup
_{v \in[0,u]} \bigl|D (u,v,f) \bigr|,
\end{equation}
where
%
\begin{equation}
\label{dfkt} D(u,v,f) = \int_0^v \mathbb{E}
\bigl[f\bigl(X_t(w)\bigr)\bigr] \,dw - \biggl(\frac
{v}{u} \biggr)
\int_0^u \mathbb{E}\bigl[f
\bigl(X_t(w)\bigr)\bigr] \,dw.
\end{equation}
If the moment function $\mathbb{E}[f(X_t(\cdot))]$ is constant on the interval
$[0,u]$, then the average $v^{-1} \int_0^v \mathbb{E}[f(X_t(w))] \,dw$
takes the
same value at all points $v \in[0,u]$, implying that $D(u,v,f) = 0$
for any $v \in[0,u]$. Hence if the function $\mathbb{E}[f(X_t(\cdot
))]$ is
constant on $[0,u]$ for any $f \in\mathcal{F}$, then the measure of
time-variation satisfies $\mathscr{D}(u) = 0$. If $\mathbb
{E}[f(X_t(\cdot))]$
varies on $[0,u]$ for some $f$ in contrast, then the average $v^{-1}
\int_0^v \mathbb{E}[f(X_t(w))] \,dw$ varies on this time span as well.
This is
ensured by the fact that $\mathbb{E}[f(X_t(\cdot))]$ is a Lipschitz continuous
function of rescaled time; that is, $|\mathbb{E}[f(X_t(w))] - \mathbb{E}
[f(X_t(w^\prime))]| \le C|w-w^\prime|$ for any $w,w^\prime\in[0,1]$,
which is a direct consequence of \eqref{gen-measure-1}. We thus obtain
that $D(u,v,f) > 0$ for some $v \in[0,u]$, which in turn yields
that $\mathscr{D}(u) > 0$. As a result, $\mathscr{D}$ satisfies
(\ref{PD}).

Since the feature $\lambda$ is constant on $[0,u_0]$ but varies after
$u_0$, the property (\ref{PD}) immediately implies that
$\mathscr{D}(u) =
0$ for $u \le u_0 $ and $\mathscr{D}(u) > 0$ for $u > u_0$. The point $u_0$
is thus characterized as the time point where the measure of
time-variation starts to deviate from zero. Importantly, the measure
$\mathscr{D}$ does not have a jump at $u_0$, but smoothly deviates
from zero
at this point. Its degree of smoothness depends on how smoothly the
moments $\mathbb{E}[f(X_t(w))]$ vary over time, or put differently, on how
smoothly the feature $\lambda(w)$ varies over time. In particular, the
smoother the time-variation in $\lambda$, the smoother the function
$\mathscr{D}$.

In order to estimate the measure of time-variation, we proceed as
follows: the integral $\int_0^v \mathbb{E}[f(X_t(w))] \,dw$ can be
regarded as
an average of the moments $\mathbb{E}[f(X_t(w))]$, where all time
points from
$0$ to $v$ are taken into account. This suggests to estimate it by the
sample average $T^{-1} \sum_{t=1}^{\lfloor vT \rfloor}
f(X_{t,T})$. Following this idea, an estimator of $\mathscr{D}(u)$ is
given by
\[
\hat{\mathscr{D}}_T(u) = \sup_{f \in\mathcal{F}} \sup
_{v \in
[0,u]} \bigl|\hat{D}_T(u,v,f) \bigr|,
\]
where
\[
\hat{D}_T(u,v,f) = \frac{1}{T} \sum
_{t = 1}^{\lfloor vT
\rfloor} f(X_{t,T}) - \biggl(
\frac{v}{u} \biggr) \frac{1}{T} \sum_{t = 1}^{\lfloor uT \rfloor}
f(X_{t,T}).
\]
The statistic $\hat{\mathscr{D}}_T(u)$ is constructed by the CUSUM principle
for the interval $[0,u]$ and can be regarded as a generalization of
classical CUSUM statistics to be found, for example, in \citeauthor{page1954} (\citeyear{page1954,page1955}). The quantity $\hat{D}_T(u,v,f)$ compares
cumulative sums of the variables $f(X_{t,T})$ over different time spans
$[0,v]$ and $[0,u]$. By taking the supremum with respect to $v \in
[0,u]$, we can detect gradual changes in the signal $\mathbb
{E}[f(X_t(\cdot)]$
on the interval $[0,u]$. The additional supremum over $f$ ensures that
the signals corresponding to all functions $f \in\mathcal{F}$ are
taken into account.

\section{Estimating the gradual change point $u_0$}\label{sec-est}

We now describe how to use our measure of time-variation to estimate
the point $u_0$. Our estimation method is based on the observation that
$\sqrt{T} \mathscr{D}(u) = 0$ for $u \le u_0 $ and $\sqrt{T}
\mathscr{D}(u)
\rightarrow\infty$ for $u > u_0$ as $T \rightarrow\infty$. The scaled
estimator $\sqrt{T} \hat{\mathscr{D}}_T(u)$ behaves in a similar
way: as we
will see later on,
%
\begin{equation}
\label{Dsup-conv}
\sqrt{T} \hat{\mathscr{D}}_T(u)
\cases{ \ds
\stackrel{d} {\longrightarrow}\mathscr{H}(u), & $\quad\mbox{for } u \le
u_0$, \vspace*{3pt}
\cr
\ds\stackrel{P} {\longrightarrow}\infty, & $
\quad\mbox{for } u > u_0$,}
\end{equation}
where $\mathscr{H}(u)$ is a real-valued random variable. By \eqref
{Dsup-conv}, $\sqrt{T} \hat{\mathscr{D}}_T(u)$ can be regarded as a statistic
to test the hypothesis that the feature of interest $\lambda$ is
time-invariant on the interval $[0,u]$. Under the null of
time-invariance, that is, as long as $u \le u_0$, the statistic weakly
converges to some limit distribution. Under the alternative, that is,
at time points $u > u_0$, it diverges in probability to infinity. The
main idea of the new estimation method is to exploit this dichotomous behavior.

To construct our estimator of $u_0$, we proceed as follows: first of
all, we define the quantity
\[
\hat{r}_T(u) = 1\bigl(\sqrt{T} \hat{\mathscr{D}}_T(u)
\le\tau_T\bigr),
\]
where $\tau_T$ is a threshold level that slowly diverges to infinity. A
data driven choice of $\tau_T$ with good theoretical and practical
properties is discussed in Sections~\ref{subsec-asym-threshold} and~\ref{sec-impl}. The random variable $\hat{r}_T(u)$ specifies the outcome of
our test on time-invariance for the interval $[0,u]$ given the critical
value $\tau_T$: if the test accepts the null of time-invariance, then
$\hat{r}_T(u) = 1$; if it rejects the null, then $\hat{r}_T(u) = 0$.
Under the null, the test statistic tends to take moderate values,
suggesting that $\hat{r}_T(u)$ should eventually become one. Under the
alternative, the statistic explodes, implying that $\hat{r}_T(u)$
should finally take the value zero. Formally speaking, one can show that
\[
\hat{r}_T(u) \stackrel{P} {\longrightarrow} %
\cases{ 1, & $\quad \mbox{for } u \le u_0$, \vspace*{3pt}
\cr
0, & $\quad\mbox{for }
u > u_0$,}
\]
if $\tau_T$ converges (slowly) to infinity. This suggests that $\int_0^1 \hat{r}_T(u) \,du \approx u_0$ for large sample sizes. Hence we may
simply estimate $u_0$ by aggregating the test outcomes $\hat{r}_T(u)$,
that is,
\[
\hat{u}_0(\tau_T) = \int_0^1
\hat{r}_T(u) \,du.
\]

\section{Asymptotic properties}\label{sec-asym}
\setcounter{equation}{0}

We now examine the asymptotic properties of the proposed estimation
method. We first investigate the weak convergence behavior of the
statistic $\hat{D}_T$ and then derive the convergence rate of the
estimator $\hat{u}_0(\tau_T)$. Due to space constraints, the proofs are
provided in the supplement [\citet{vogtdett2014}]. To state the results,
we let the symbol $\ell_{\infty}(S)$ denote the space of bounded
functions $f\dvtx  S \rightarrow\mathbb{R}$ endowed with the supremum norm and
let $\rightsquigarrow$ denote weak convergence. Moreover, to capture
the amount
of smoothness of the measure $\mathscr{D}$ at the point $u_0$, we
suppose that
%
\begin{equation}
\label{smoothness-D}
\frac{\mathscr{D}(u)}{(u - u_0)^{\kappa}} \rightarrow c_{\kappa} > 0 \qquad \mbox{as
} u \searrow u_0
\end{equation}
for some number $\kappa> 0$ and a constant $c_{\kappa} > 0$. The
larger $\kappa$, the more smoothly the measure $\mathscr{D}$ deviates from
zero at the point $u_0$.

\subsection{Assumptions}\label{subsec-asym-ass}

Throughout the paper, we make the following assumptions:
\begin{longlist}[(C3)]
\item[(C1)] The process $\{X_{t,T}\}$ is locally stationary in the
sense of Definition~\ref{def-loc-stat}.
\item[(C2)] The process $\{X_{t,T}\}$ is strongly mixing with
mixing coefficients $\alpha(k)$ satisfying $\alpha(k) \le C a^k$ for
some positive constants $C$ and $a < 1$.
\item[(C3)] Let $p \ge4$ be an even natural number, and endow the
set $\mathcal{F}$ with some semimetric $d_{\mathcal{F}}$. $(\mathcal
{F},d_{\mathcal{F}})$
is separable, compact and not too complex in the sense that its
covering number $\mathcal{N}(w,\mathcal{F},d_{\mathcal{F}})$
satisfies the condition
$\int_0^1 \mathcal{N}(w,\mathcal{F},d_{\mathcal{F}})^{1/p} \,dw <
\infty$.
Moreover, the set $\mathcal{F}$ has an envelope $F$ (i.e., $|f| \le F$
for all $f \in\mathcal{F}$) which satisfies $\mathbb
{E}[F(X_{t,T})^{(1+\delta
)p}] \le C < \infty$ for some small $\delta> 0$ and a fixed constant
$C$. Finally, for any pair of functions $f,f^{\prime} \in\mathcal{F}$,
\[
\mathbb{E} \biggl[ \biggl| \frac{f(X_{t,T}) - f^{\prime
}(X_{t,T})}{d_{\mathcal{F}}
(f,f^{\prime})} \biggr|^{(1+\delta)p} \biggr] \le C < \infty.
\]
\item[(C4)] For $k=1,2$ and all $f \in\mathcal{F}$, it holds that
$\mathbb{E}[|f(X_{t,T}) - f(X_t(u))|^k] \le C (|\frac{t}{T}-u| +
\frac
{1}{T})$ for some fixed constant $C$.
\end{longlist}

Finally, we take for granted that all suprema over $\mathcal{F}$ which show up in the
course of our analysis are measurable.

(C2) stipulates that the array $\{X_{t,T}\}$ is strongly mixing. A
wide variety of locally stationary processes are mixing under
appropriate conditions; see, for example, \citet{SubbaRao2011} and
\citet{Vogt2012}. To keep the structure of the proofs as clear as possible,
we have assumed the mixing rates to decay exponentially fast.
Alternatively, we could work with slower polynomial rates at the cost
of a more involved notation in the proofs. (C3) and (C4) allow
for a wide range of function families $\mathcal{F}$ and are formulated
in a very general way. For many choices of $\mathcal{F}$, they boil
down to simple moment conditions on the variables $X_{t,T}$, $X_t(u)$
and $U_{t,T}(u)$. We illustrate this by means of Examples \ref{example1}--\ref{example3}. It is straightforward to show that in
Example~\ref{example1}, (C3) and (C4) are satisfied under the
following set of moment conditions:
\begin{longlist}[$(C_{\mu})$]
\item[$(C_{\mu})$] Either (a) $\mathbb{E}|X_{t,T}|^r \le C$ for some
$r > 4$
and $\mathbb{E}U_{t,T}^2(u) \le C$ or (b) $\mathbb{E}|X_{t,T}|^r \le
C$, $\mathbb{E}
|X_t(u)|^r \le C$ and $\mathbb{E}U_{t,T}^{r/(r-1)}(u) \le C$ for some
$r > 4$
and a sufficiently large constant $C$ that is independent of $u$, $t$
and $T$.
\end{longlist}
Similarly, in Example~\ref{example2}, they are implied by
\begin{longlist}[$(C_{\gamma})$]
\item[$(C_{\gamma})$] $\mathbb{E}\|X_{t,T}\|^r \le C$, $\mathbb{E}\|
X_t(u)\|^r \le
C$ and $\mathbb{E}U_{t,T}^q(u) \le C$ for some $r > 8$ and $q = \frac
{r}{3} /
(\frac{r}{3}-1)$, where $C$ is a sufficiently large constant that is
independent of $u$, $t$ and $T$.
\end{longlist}
The moment conditions in Example~\ref{example3} are fully analogous to
those in Example~\ref{example2} and thus not stated explicitly.
Finally, note that our conditions (C1)--(C4) allow for families
$\mathcal{F}$ of nonsmooth functions in principle. However, it is
technically more difficult to verify (C3) and (C4) for such
families than for classes of smooth functions.

\subsection{Weak convergence of the measure of time-variation}\label
{subsec-asym-measure}

To start with, we investigate the asymptotic properties of the expression
%
\begin{equation}
\label{Hpt} \hat{H}_T(u,v,f) = \sqrt{T} \bigl(
\hat{D}_T(u,v,f) - D(u,v,f) \bigr).
\end{equation}
To do so, let $\Delta= \{(u,v) \in[0,1]^2\dvtx  v \le u \}$, and equip the
space $\Delta\times\mathcal{F}$ with the natural semimetric $|u -
u^\prime| + |v - v^\prime| + d_{\mathcal{F}}(f,f^\prime)$. In what
follows, we
regard $\hat{H}_T$ as a process that takes values in $\ell_{\infty
}(\Delta\times\mathcal{F})$ and show that it weakly converges to a
Gaussian process $H$ with the covariance structure
%
\begin{eqnarray}
&&\operatorname{Cov}\bigl(  H(u,v,f),H\bigl(u^\prime,v^\prime,f^\prime
\bigr)\bigr)\nonumber
\\
\label{cov-H}
&&\qquad= \sum_{l=-\infty}^{\infty} \biggl\{
\frac{v v^\prime}{u
u^\prime} \int_0^{\min\{u,u^\prime\}}
c_l(w) \,dw - \frac{v^\prime
}{u^\prime} \int_0^{\min\{v,u^\prime\}}
c_l(w) \,dw
\\
&&\hspace*{49pt}\qquad {}- \frac{v}{u} \int_0^{\min\{u,v^\prime\}}
c_l(w) \,dw + \int_0^{\min\{v,v^\prime\}}
c_l(w) \,dw \biggr\},
\nonumber
\end{eqnarray}
where $c_l(w) = c_l(w,f,f^{\prime}) = \operatorname
{Cov}(f(X_0(w)),f^\prime
(X_l(w)))$. The following theorem gives a precise description of the
weak convergence of $\hat{H}_T$.

\begin{theorem}\label{theo-measure}
Let assumptions \textup{(C1)}--\textup{(C4)} be satisfied. Then
\[
\hat{H}_T = \sqrt{T} [ \hat{D}_T - D ] \rightsquigarrow
H
\]
as a process in $\ell_{\infty}(\Delta\times\mathcal{F})$, where
$\hat
{D}_T$ and $D$ are defined in Section~\ref{sec-measure}, and $H$
is a centered Gaussian process on $\Delta\times\mathcal{F}$ with
covariance kernel \eqref{cov-H}.
\end{theorem}

This theorem is the main stepping stone to derive the asymptotic
properties of our estimator $\hat{u}_0(\tau_T)$. In addition, it is
useful to examine the asymptotic behavior of some processes related to
$\hat{H}_T$: analogously to $\hat{\mathscr{D}}_T(u)$, we introduce
the expression
%
\begin{equation}
\label{Hsup} \hat{\mathscr{H}}_T(u) = \sup_{f \in\mathcal{F}}
\sup_{v \in
[0,u]} \bigl| \hat{H}_T(u,v,f) \bigr|.
\end{equation}
Moreover, we let
%
\begin{eqnarray}\label{Dmax}
\hat{\mathbb{D}}_T(u) & =& \sup_{v \in[0,u]} \hat{\mathscr
{D}}_T(v) = \sup_{f \in
\mathcal{F}} \sup_{0 \le w \le v \le u}
\bigl| \hat{D}_T(v,w,f) \bigr|,
\\
\label{Hmax}
\hat{\mathbb{H}}_T(u) & =& \sup_{v \in[0,u]} \hat{\mathscr
{H}}_T(v) = \sup_{f \in
\mathcal{F}} \sup_{0 \le w \le v \le u}
\bigl| \hat{H}_T(v,w,f) \bigr|.
\end{eqnarray}
The next result directly follows from Theorem~\ref{theo-measure}
together with the continuous mapping theorem.

\begin{corollary}\label{corollary-measure}
Let assumptions \textup{(C1)}--\textup{(C4)} be satisfied. Then $\hat{\mathscr{H}}_T
\rightsquigarrow\mathscr{H}$ and $\hat{\mathbb{H}}_T
\rightsquigarrow\mathbb{H}$ as processes in $\ell
_{\infty}([0,1])$, where $\mathscr{H}$ and $\mathbb{H}$ are defined
by $\mathscr{H}(u) =
\sup_{f \in\mathcal{F}, v \in[0,u]} |H(u,v,f)|$ and $\mathbb{H}(u) =
\sup_{f \in\mathcal{F}, 0 \le w \le v \le u} |H(v,w,f)|$, respectively.
\end{corollary}

\subsection{Convergence of the estimator \texorpdfstring{$\hat{u}_0(\tau_T)$}{hat{u}0(tau_T)}}\label
{subsec-asym-convergence}

We now turn to the asymptotic behavior of the estimator $\hat
{u}_0(\tau
_T)$. The next theorem shows that $\hat{u}_0(\tau_T)$ consistently
estimates $u_0$ provided that the threshold level $\tau_T$ diverges to
infinity.

\begin{theorem}\label{theo-convergence}
Let assumptions \textup{(C1)}--\textup{(C4)} be satisfied and assume that $\tau_T
\rightarrow\infty$ with $\tau_T / \sqrt{T} \rightarrow0$. Then
\[
\hat{u}_0(\tau_T) - u_0 =
O_p(\gamma_T),
\]
where $\gamma_T = (\tau_T / \sqrt{T})^{1/\kappa}$, and $\kappa$ is
defined in \eqref{smoothness-D}.
\end{theorem}

The convergence rate $\gamma_T$ of $\hat{u}_0(\tau_T)$ depends on the
degree of smoothness $\kappa$ of the measure $\mathscr{D}$ at the point
$u_0$. In particular, when $\mathscr{D}$ is smoother, the convergence
rate is
slower. Since the smoothness of $\mathscr{D}$ mirrors that of the stochastic
feature $\lambda$, we can equivalently say: the smoother the feature
$\lambda$ varies around $u_0$, the slower the rate of our estimator.
This reflects the intuition that it becomes harder to precisely
localize the point $u_0$ when $\lambda$ varies more slowly and smoothly
around this point. The rate $\gamma_T$ also depends on the threshold
parameter $\tau_T$. Specifically, the more slowly $\tau_T$ diverges to
infinity, the faster $\gamma_T$ goes to zero.

\subsection{Some mean squared error considerations}

A more detailed picture of the convergence behavior of $\hat{u}_0(\tau
_T)$ can be obtained by examining the mean squared error $\mbox
{MSE}(\tau_T) = \mathbb{E}[ (\hat{u}_0(\tau_T) - u_0)^2]$. Let us
first ignore
the estimation error in $\hat{\mathscr{D}}_T$ and suppose that $\hat
{\mathscr{D}}_T
= \mathscr{D}$. In this noiseless case, the MSE has the form
%
\begin{equation}
\label{MSE-1}
\operatorname{MSE}(\tau_T) = \biggl\{ \int
_{u_0}^1 1\bigl(\sqrt{T} \mathscr {D}(u) \le\tau
_T\bigr) \,du \biggr\}^2.
\end{equation}
Under our smoothness condition \eqref{smoothness-D}, it is not
difficult to see that
%
\begin{equation}
\label{MSE-2}
C_1 \gamma_T^2 \le\operatorname{MSE}(
\tau_T) \le C_2 \gamma_T^2
\end{equation}
for some constants $0 < C_1 \le C_2 < \infty$. Inequality \eqref{MSE-2}
can be refined if we replace~\eqref{smoothness-D} by a more specific
smoothness condition. In particular, suppose that $\mathscr{D}$ has a
cusp of
order $\kappa$ at $u_0$ in the following sense: (i) it admits the
Taylor-type expansion
\[
\mathscr{D}(u) = \mathscr{D}(u_0) + \sum
_{k = 1}^{\kappa} \frac{\mathscr{D}
^{(k)}(u_0)}{k!} (u -
u_0)^k + R(u)
\]
for $u \in[u_0,u_0+\delta]$, where $\mathscr{D}^{(k)}(u_0)$ is the $k$th
order right derivative of $\mathscr{D}$ at $u_0$ and the remainder $R(u)$
satisfies $|R(u)| \le C (u - u_0)^{\kappa+1}$, and (ii) $\mathscr{D}
^{(k)}(u_0) = 0$ for all $1 \le k \le\kappa- 1$ but $\mathscr
{D}^{(\kappa
)}(u_0) \ne0$. This implies that $\mathscr{D}(u) = \frac{\mathscr
{D}^{(\kappa
)}(u_0)}{\kappa!} (u - u_0)^{\kappa} + R(u)$ for $u \in
[u_0,u_0+\delta
]$, which in turn yields that for any $\varepsilon> 0$,
%
\begin{equation}
\label{MSE-3} (1 - \varepsilon) \frac{\mathscr{D}^{(\kappa)}(u_0)}{\kappa!} (u - u_0)^{\kappa}
\le\mathscr{D}(u) \le(1 + \varepsilon) \frac
{\mathscr{D}^{(\kappa
)}(u_0)}{\kappa!} (u -
u_0)^{\kappa}
\end{equation}
as $u \searrow u_0$. Plugging \eqref{MSE-3} into \eqref{MSE-1}, we can
infer that
%
\begin{equation}\label{MSE-4}
\quad \biggl(\frac{\kappa!}{(1+\varepsilon)\mathscr{D}^{(\kappa
)}(u_0)} \biggr)^{2/\kappa
}
\gamma_T^2 \le\operatorname{MSE}(\tau_T) \le \biggl(
\frac{\kappa
!}{(1-\varepsilon)\mathscr{D}^{(\kappa)}(u_0)} \biggr)^{2/\kappa} \gamma_T^2
\end{equation}
for $T$ large enough. \eqref{MSE-4} is a refinement of \eqref{MSE-2},
which shows quite explicitly how the MSE depends on the smoothness of
$\mathscr{D}$, and in particular, how it depends on the higher-order right
derivatives of $\mathscr{D}$ at $u_0$.

We next compute the MSE in the noisy case where $\hat{\mathscr{D}}_T$
involves some estimation error. Using the shorthand
\[
\vartheta_T = \biggl\{ \int_0^{u_0}
\bigl[ 1\bigl(\sqrt{T} \hat{\mathscr {D}}_T(u) \le \tau_T
\bigr) - 1 \bigr] \,du + \int_{u_0}^1 1\bigl(
\sqrt{T} \hat{\mathscr {D}}_T(u) \le\tau _T\bigr) \,du
\biggr\}^2,
\]
we can write $\operatorname{MSE}(\tau_T) = \operatorname{MSE}_1(\tau_T) + \mbox
{MSE}_2(\tau_T)$, where
\begin{eqnarray*}
\operatorname{MSE}_1(\tau_T) & =& \mathbb{E} \bigl[
\vartheta_T 1 \bigl(\hat {\mathbb{H}}_T(1) \le
b_T\bigr) \bigr],
\\
\operatorname{MSE}_2(\tau_T) & =& \mathbb{E} \bigl[
\vartheta_T 1 \bigl(\hat {\mathbb{H}}_T(1) >
b_T\bigr) \bigr]
\end{eqnarray*}
and $b_T \rightarrow\infty$ with $b_T/\tau_T \rightarrow0$. Noting
that the estimation error $|\sqrt{T} \hat{\mathscr{D}}_T(u) - \sqrt
{T} \mathscr{D}
(u)|$ is bounded by $\hat{\mathbb{H}}_T(1)$, the two terms $\operatorname{MSE}_1(\tau
_T)$ and $\operatorname{MSE}_2(\tau_T)$ can be interpreted as follows: $\operatorname{MSE}_1(\tau_T)$ represents the case where the estimation error is
small (in particular smaller than $b_T$) and can be effectively
neglected. As shown in the supplement [\citet{vogtdett2014}],
%
\begin{equation}
\label{MSE-5} C_1 \gamma_T^2 \le
\operatorname{MSE}_1(\tau_T) \le C_2
\gamma_T^2
\end{equation}
under condition \eqref{smoothness-D}, where the constants $C_1$ and
$C_2$ can be chosen as in \eqref{MSE-2}. In addition, it is possible to
derive an inequality of the form \eqref{MSE-4} under the strengthened
smoothness condition. Hence, $\operatorname{MSE}_1(\tau_T)$ essentially behaves
as the MSE in the noiseless case. $\operatorname{MSE}_2(\tau_T)$ represents the
situation where the estimation error is large (in particular, larger
than $b_T$). It thus captures the main effect of the estimation error
on the MSE. A rough bound is
\[
\operatorname{MSE}_2(\tau_T) \le\mathbb{P} \bigl( \hat{
\mathbb{H}}_T(1) > b_T \bigr).
\]
This yields $\operatorname{MSE}(\tau_T) \le O(\gamma_T^2) + \mathbb{P}(\hat
{\mathbb{H}
}_T(1) > b_T)$, which approximately separates the noiseless part of the
MSE from the effect of the estimation error.

Similarly to bandwidth selection for kernel estimators, we may want to
choose the tuning parameter $\tau_T$ by minimizing the MSE with respect
to it. Even though appealing, this is extremely difficult to achieve.
The main problem is to derive an exact asymptotic expression for $\mbox
{MSE}_2(\tau_T)$: the asymptotic properties of this term strongly
depend on the tail behavior of the estimation error $\sqrt{T} (\hat
{\mathscr{D}}_T - \mathscr{D})$. To compute an exact asymptotic
formula, we thus
need a precise description of the tail behavior of the estimation
error, which seems intractable in general. For this reason, we propose
an alternative procedure for the choice of $\tau_T$ in what follows,
which is similar to the choice of the critical value in a test procedure.

\subsection{Choice of the threshold level \texorpdfstring{$\tau_T$}{tauT}}\label
{subsec-asym-threshold}

We now discuss how to choose the threshold $\tau_T$ to obtain an
estimator of $u_0$ with good theoretical properties. To state the
results, we let $q_{\alpha}(u)$ be the $(1-\alpha)$-quantile of the
limit variable $\mathbb{H}(u)$ and assume throughout that this
quantile is
known for any time point $u$. In practice, it is indeed unknown and has
to be approximated. We show how to achieve this in Section~\ref{sec-impl} where we discuss the implementation of our method. Our
choice of the threshold $\tau_T$ proceeds in two steps. In the first,
we describe a rough choice of $\tau_T$ which leads to a preliminary
estimator of $u_0$. In the second, we use this preliminary estimator to
come up with a refined choice of $\tau_T$, which in turn yields a
better estimator of $u_0$.

\subsubsection*{Preliminary choice of \texorpdfstring{$\tau_T$}{tauT}}
To convey the idea behind the
choice of $\tau_T$, let us first assume that $\tau_T$ does not depend
on the sample size, that is, $\tau_T = \tau$ for all $T$. A first crude
choice of $\tau$ can be obtained by arguing in a similar way as in
classical change point detection problems: consider the situation that
the stochastic feature of interest is time-invariant on $[0,1]$; that
is, there is no change point $u_0 < 1$. In this situation, we would
like to control the probability of false detection of a change point.
Specifically, we aim to choose $\tau$ such that this probability is
smaller than some pre-specified level $\alpha$, that is,
\[
\mathbb{P}\bigl( \hat{u}_0(\tau) < 1\bigr) \le\alpha,
\]
when there is no change point $u_0 < 1$. To achieve this, we write
\[
\mathbb{P}\bigl(\hat{u}_0(\tau) < 1\bigr)  \le\mathbb{P} \bigl(
\sqrt{T} \hat{\mathscr{D}}_T(u) > \tau \mbox{ for some } u \in[0,1]
\bigr) = \mathbb{P} \bigl( \sqrt{T} \hat{\mathbb{D}}_T(1) > \tau
\bigr).
\]
Corollary~\ref{corollary-measure} shows that $\sqrt{T} \hat{\mathbb{D}
}_T(u)$ weakly converges to the limit variable $\mathbb{H}(u)$ at each point
$u \le u_0$. In particular, when there is no change point $u_0 < 1$,
then $\sqrt{T} \hat{\mathbb{D}}_T(1) \stackrel{d}{\longrightarrow
}\mathbb{H}(1)$. We now set $\tau$ to
be the $(1-\alpha)$-quantile $q_{\alpha}(1)$ of $\mathbb{H}(1)$. Writing
$\tau_{\alpha}^{\circ} = q_{\alpha}(1)$, we obtain that
\[
\mathbb{P}\bigl(\hat{u}_0\bigl(\tau_{\alpha}^{\circ}
\bigr) < 1\bigr) \le\alpha+ o(1),
\]
when there is no change point $u_0 < 1$. We are thus able to
asymptotically control the probability of false detection by choosing
$\tau= \tau_{\alpha}^{\circ}$.

Of course, this choice does not yield a consistent estimator of $u_0$.
To ensure consistency, we have to make sure that $\tau_T$ diverges to
infinity. To achieve this, we let the level $\alpha_T$ depend on $T$.
In particular, we let it converge to zero and set $\tau_T = \tau
_{\alpha
_T}^{\circ}$. Using Borell's inequality, we can show that $\mathbb{H}(1)$
has exponentially decaying tails and infer that $\tau_{\alpha
_T}^{\circ
}/\sqrt{T} \rightarrow0$ as long as $\alpha_T \ge cT^{-r}$ for some
positive constants $c$ and $r$. Hence, as long as $\alpha_T$ converges
to zero at polynomial speed at most, the choice $\tau_T = \tau
_{\alpha
_T}^{\circ}$ satisfies the requirements of Theorem~\ref
{theo-convergence} and thus results in a consistent estimator of $u_0$.

\subsubsection*{Refined choice of \texorpdfstring{$\tau_T$}{tauT}}
As in classical change point
problems, the choice $\tau= \tau_{\alpha}^{\circ}$ is fairly
conservative. In particular, the resulting estimator tends to strongly
overestimate the time point $u_0$. In what follows, we refine the
choice of $\tau$ to get a more precise estimator of $u_0$. Rather than
controlling the false detection rate, we would like to control the
probability of underestimating $u_0$, that is, of falsely detecting a
change point before it occurs. Technically speaking, we aim to choose
$\tau$ such that
\[
\mathbb{P}\bigl(\hat{u}_0(\tau) < u_0\bigr) \le\alpha
\]
for some given level $\alpha$. Similarly as above, it holds that
\[
\mathbb{P}\bigl(\hat{u}_0(\tau) < u_0\bigr)  \le
\mathbb{P} \bigl( \sqrt{T} \hat{\mathscr{D}}_T(u) > \tau \mbox{ for
some } u \in[0,u_0] \bigr) = \mathbb{P} \bigl( \sqrt{T} \hat{
\mathbb{D}}_T(u_0) > \tau\bigr).
\]
By Corollary~\ref{corollary-measure}, we know that $\sqrt{T} \hat
{\mathbb{D}
}_T(u_0) \stackrel{d}{\longrightarrow}\mathbb{H}(u_0)$. Setting
$\tau$ to equal the $(1-\alpha
)$-quantile $q_{\alpha}(u_0)$ of $\mathbb{H}(u_0)$ and writing $\tau
_{\alpha
} = q_{\alpha}(u_0)$, we are able to derive the following result.

\begin{theorem}\label{theo-threshold}
Under assumptions \textup{(C1)}--\textup{(C4)}, it holds that
%
\begin{equation}
\label{error-underest}
\mathbb{P}\bigl(\hat{u}_0(\tau_{\alpha}) <
u_0\bigr) \le\alpha+ o(1).
\end{equation}
Moreover, for any constant $C > 0$ and $\gamma_T$ as defined in Theorem~\ref{theo-convergence},
%
\begin{equation}
\label{error-overest}
\mathbb{P}\bigl(\hat{u}_0(\tau_{\alpha}) >
u_0 + C \gamma_T\bigr) = o(1).
\end{equation}
\end{theorem}

Hence the estimator $\hat{u}_0(\tau_{\alpha})$ has the following
properties: according to \eqref{error-underest}, the~probability of
underestimating $u_0$ is asymptotically bounded by $\alpha$. Moreover,
the probability of overestimating $u_0$ by more than $C \gamma_T$ is
asymptotically negligible by \eqref{error-overest}. Thus $\hat
{u}_0(\tau
_{\alpha})$ controls the error of underestimating $u_0$ while being
consistent when it comes to overestimation.

Of course, we cannot take the choice $\tau= \tau_{\alpha}$ at face
value since the quantile $\tau_{\alpha} = q_{\alpha}(u_0)$ depends on
the unknown location $u_0$. Nevertheless, we can estimate this quantile
by $\hat{\tau}_{\alpha} = q_{\alpha}(\hat{u}_0^\circ)$, where
$\hat
{u}_0^{\circ}$ is a consistent pilot estimator of $u_0$. In particular,
we may set $\hat{u}_0^\circ= \hat{u}_0(\tau_{\alpha_T}^{\circ})$. It\vspace*{1pt}
is fairly straightforward to see that the statements of Theorem~\ref
{theo-threshold} still hold true when $\tau_{\alpha}$ is replaced by
$\hat{\tau}_{\alpha}$:

\begin{corollary}\label{corollary-threshold}
Let assumptions \textup{(C1)}--\textup{(C4)} be satisfied. Then
%
\begin{equation}
\label{error-underest-cor} \mathbb{P}\bigl(\hat{u}_0(\hat{\tau}_{\alpha})
< u_0\bigr) \le\alpha+ o(1)
\end{equation}
and for any $C > 0$,
%
\begin{equation}
\label{error-overest-cor} \mathbb{P}\bigl(\hat{u}_0(\hat{\tau}_{\alpha})
> u_0 + C \gamma_T\bigr) = o(1).
\end{equation}
\end{corollary}

To obtain a consistent estimator of $u_0$, we suggest setting $\tau_T =
\hat{\tau}_{\alpha_T}$ with $\alpha_T$ gradually converging to
zero. As
above, we can ensure consistency as long as $\alpha_T$ converges to
zero at polynomial rate at most.

\section{Implementation}\label{sec-impl}

In practice, we implement our method as\vspace*{-2pt} follows:
\begin{enumerate}[\textit{Step} 2]

\item[\textit{Step} 1.] Choose the tuning parameter $\tau_{\alpha}$:
\begin{enumerate}[(b)]
\item[(a)] Fix a probability level $\alpha$, and estimate the quantiles
$q_{\alpha}(u)$ by $\hat{q}_{\alpha}(u)$ as described below.
\item[(b)] Compute the preliminary estimator $\hat{u}_0^{\circ} = \hat
{u}_0(\hat{\tau}_{\alpha}^{\circ})$, where $\hat{\tau}_{\alpha
}^{\circ}
= \hat{q}_{\alpha}(1)$.
\item[(c)] Estimate $\tau_{\alpha}$ by $\hat{\tau}_{\alpha} = \hat
{q}_{\alpha
}(\hat{u}_0^{\circ})$.
\end{enumerate}
\item[\textit{Step} 2.] Estimate $u_0$ by $\hat{u}_0(\hat{\tau
}_{\alpha})$.
\end{enumerate}
To complete the description of our implementation strategy, we explain
how to estimate the quantiles $q_{\alpha}(u)$ of $\mathbb{H}(u)$. We first
set up a general estimator and then explain how to simplify it in some
special settings that are of particular importance in practice.\vspace*{-3pt}

\subsection*{The general case}
By definition, $\mathbb{H}(u)$ is a
supremum of
the Gaussian process $H$ whose covariance structure is given in
\eqref{cov-H}. Inspecting formula \eqref{cov-H}, the only unknown
expressions occurring in it are of the\vspace*{-2pt} form
\[
\sigma^2\bigl(u,f,f^{\prime}\bigr) = \sum
_{l=-\infty}^{\infty} \Gamma _l
\bigl(u,f,f^{\prime}\bigr),
\]
where $\Gamma_l(u,f,f^{\prime}) = \int_0^u c_l(w,f,f^{\prime}) \,dw$ and
$c_l(w,f,f^{\prime}) = \operatorname{Cov}(f(X_0(w)),\break  f^{\prime
}(X_l(w)))$. These
quantities are essentially average long-term covariances of the
processes $\{f(X_t(w))\}$ and $\{f^{\prime}(X_t(w))\}$ on the interval
$[0,u]$, which can be estimated by methods for long-run covariance
estimation. Specifically, we can employ HAC-type estimation procedures,
as discussed in \citet{Andrews1991} or \citet{deJong2000}, and work with
an estimator of the form
%
\begin{equation}
\label{est-lrv} \hat{\sigma}^2\bigl(u,f,f^{\prime}\bigr) = \sum
_{l=-\infty}^{\infty
} W \biggl(\frac{l}{b(T)}
\biggr) \hat{\Gamma}_l\bigl(u,f,f^{\prime}\bigr),
\end{equation}
where $W$ is a kernel of Bartlett or flat-top type and $b=b(T)$ is the
bandwidth. Moreover,\vspace*{-6pt}
\[
\hat{\Gamma}_l\bigl(u,f,f^{\prime}\bigr) = \frac{1}{T}
\sum_{t=1}^{\lfloor
uT \rfloor} \hat{Z}_{t,T}(f)
\hat{Z}_{t-l,T}\bigl(f^{\prime}\bigr),
\]
where $\hat{Z}_{t,T}(f) = f(X_{t,T}) - \hat{\mathbb{E}}[f(X_{t,T})]$
and $\hat
{\mathbb{E}}[f(X_{t,T})]$ is an estimator of $\mathbb
{E}[f(X_{t,T})]$. Specifically,
we may work with the Nadaraya--Watson estimator $\hat{\mathbb{E}}[f(X_{t,T})]
= \sum_{s=1}^T K_h(\frac{t}{T} - \frac{s}{T}) f(X_{s,T})/ \sum_{s=1}^T K_h(\frac{t}{T} - \frac{s}{T})$, where $h$ is the bandwidth
and $K$ is a kernel function with $K_h(x) = h^{-1} K(x/h)$.
Alternatively, a local linear smoother may be employed.

Once we have calculated $\hat{\sigma}^2(u,f,f^{\prime})$, we can
approximate the covariance function $C(u,u^\prime,v,v^\prime
,f,f^\prime
) = \operatorname{Cov}(H(u,v,f),H(u^\prime,v^\prime,f^\prime))$ of
\eqref
{cov-H} by an estimator $\hat{C}(u,u^\prime,v,v^\prime,f,f^\prime)$ and
simulate observations from the Gaussian process with the estimated
covariance structure as follows: let $0 = u_1 < u_2 < \cdots< u_n = 1$
be a grid of time points and $\mathcal{F} = \{ f_1,\ldots,f_K\}$ a
(finite) family of functions. Calculate the covariance matrix $M = \{
M_{k,k^\prime}\dvtx  1 \le k,k^\prime\le K \}$, where $M_{k,k^\prime} = \{
\hat{C}(u_i,u_{i^\prime},u_j,u_{j^\prime},f_k,f_{k^\prime}) \dvtx  1 \le
j <
i \le n, 1 \le j^\prime< i^\prime\le n \}$ is the covariance matrix
that corresponds to the pair of functions $(f_k,f_{k^\prime})$. Then
simulate normally distributed random vectors with the covariance
structure $M$. (Note that the matrix $M$ is not necessarily positive
semi-definite for fixed $T$. We thus modify it as follows in practice:
let $M = S A S^{\intercal}$ be the eigendecomposition of $M$, where
$A$ is a diagonal matrix whose entries are the eigenvalues of $M$ and
$S$ is an orthonormal matrix. Now set all eigenvalues below a given
small threshold equal to zero, and denote the corresponding diagonal
matrix by $A^\prime$. We then replace $M$ by $M^\prime= S A^\prime
S^{\intercal}$, which is positive semi-definite by construction.)

The simulated normal random vectors with the covariance matrix $M$
approximate observations of the Gaussian process $H$. From these, we
can calculate approximations $\mathbb{H}_{\mathrm{approx}}(u_i)$ of the
variables $\mathbb{H}(u_i)$ and compute the empirical $(1-\alpha)$-quantile
$\hat{q}_{\alpha}(u_i)$ of $\mathbb{H}_{\mathrm{approx}}(u_i)$ for
each $i$.
These serve as estimates of the quantiles $q_{\alpha}(u_i)$ at the grid
points\vspace*{6pt} $u_i$.

Our implementation strategy works well in practice as we will
demonstrate in the empirical part of the paper. Of course, when the
class of functions $\mathcal{F}$ is large, it becomes computationally
more burdensome to simulate the quantiles $q_{\alpha}(u)$. In most
applications, however, the class of functions is fairly small.
Moreover, in a number of cases, we can modify our estimation procedure
such that the quantiles $q_{\alpha}(u)$ need not be estimated but are
known (up to simulation error). To illustrate this, we consider two settings.

\subsection*{Setting \textup{I}} We first revisit the time-varying mean setting from
Example~\ref{example1}. Specifically, consider the model\vspace*{-3pt}
%
\begin{equation}
\label{mod0} X_{t,T} = \mu \biggl( \frac{t}{T} \biggr) +
\varepsilon_{t,T},
\end{equation}
where the feature $\lambda_{t,T}$ of interest is the mean function
$\mu
(\frac{t}{T})$. Recalling that $\mathcal{F} = \{ \mbox{id} \}$ in this
case, the covariance structure \eqref{cov-H} depends on the expressions
$\sigma^2(u) = \sum_{l=-\infty}^{\infty} \int_0^u c_l(w) \,dw$,
where $c_l(w) = \mathbb{E}[\varepsilon_0(w) \varepsilon_l(w)]$ and
$\{
\varepsilon_t(w)\}$ is the stationary approximating process of $\{
\varepsilon_{t,T}\}$ at the time point $w$. If the error process is
stationary, we even obtain that $\sigma^2(u) = u \sigma^2$ for all $u$,
where $\sigma^2 = \sum_{l=-\infty}^{\infty} \mathbb{E}[
\varepsilon_0
\varepsilon_l]$ is the long-run variance of the error terms. The latter
can be estimated by standard methods. Denoting its estimator by $\hat
{\sigma}^2$, we replace the statistic $\hat{\mathscr{D}}_T(u)$ in our
estimation method by the scaled version $\hat{\mathscr{D}}_T^{\mathrm
{sc}}(u) =
\hat{\mathscr{D}}_T(u)/\hat{\sigma}$.
Defining  $\hat{H}_T^{\mathrm{sc}} = \sqrt{T}(\hat
{D}_T - D)/\hat{\sigma}$,
we obtain that $\hat{H
}_T^{\mathrm{sc}} \rightsquigarrow H^{\mathrm{sc}}$, where the Gaussian process
$H^{\mathrm{sc}}$ has the covariance structure $\operatorname
{Cov}(H^{\mathrm
{sc}}(u,v),H^{\mathrm{sc}}(u^\prime,v^\prime)) = C(u,u^\prime
,v,v^\prime)$\vspace*{-2pt} with
%
\begin{eqnarray}
C\bigl(u,u^\prime,v,v^\prime\bigr) & =& \frac{v v^\prime}{u u^\prime} \min
\bigl\{ u,u^\prime\bigr\} - \frac{v^\prime}{u^\prime} \min\bigl
\{v,u^\prime\bigr\}
\nonumber
\\[-9pt]
\label{known-cov}
\\[-9pt]
&&{}- \frac{v}{u} \min\bigl\{u,v^\prime\bigr\} + \min\bigl\{
v,v^\prime\bigr\}.
\nonumber
\end{eqnarray}
Importantly, this formula does not involve any unknown quantities,
implying that the quantiles $q_{\alpha}^{\mathrm{sc}}(u)$\vspace*{1pt} of $\mathbb
{H}^{\mathrm{sc}}(u) =\sup_{0\leq w\leq v\leq u}|H^{\mathrm{sc}}(v,w)|$ are known (up to simulation error).
Inspecting the proofs of Theorems~\ref{theo-convergence}--\ref{theo-threshold} and
Corollary~\ref{corollary-threshold}, the quantiles $q^{\mathrm{sc}}_{\alpha}(u)$ can be seen to take the place of
$q_{\alpha}(u)$
when we work with the scaled statistic
$\hat{\mathscr{D}}^{\mathrm{sc}}_{T}$.
Consequently, setting up
our estimation method in terms of $\hat{\mathscr{D}}_T^{\mathrm{sc}}$ rather
than $\hat{\mathscr{D}}_T$, we can make use of  the known quantiles
$q_{\alpha}^{\mathrm{sc}}(u)$ in Step 1 of our implementation strategy.

\subsection*{Setting \textup{II}} A similar simplification of our estimation
procedure is possible whenever (i) the class $\mathcal{F} = \{
f_1,\ldots,f_K \}$ is finite, and (ii) the process $\{ X_{t,T} \}$ is
stationary up to the time point $u_0$. It goes without saying that (i)
is fulfilled in many cases. The same holds for (ii): obviously, it is
satisfied in the mean setting~\eqref{mod0} when the error process is
stationary. More generally speaking, it is often met when $\{X_{t,T}\}$
is modeled by a process with time-varying parameters. Examples are
moving average processes of the form $X_{t,T} = \sum_{\ell
=0}^{\infty} a_{\ell}(\frac{t}{T}) \varepsilon_{t-\ell}$ or
autoregressive processes $X_{t,T} = \sum_{\ell=1}^{p}
a_{\ell
}(\frac{t}{T}) X_{t-\ell,T} + \varepsilon_t$, where $a_{\ell}$ are
time-varying parameter functions and $\varepsilon_t$ are i.i.d.
residuals. These processes are stationary as long as the model
parameters do not vary over time. In many applications, it makes sense
to assume that the parameters are constant up to some time point $u_0$
and then start to vary. Hence, in these cases, the process $\{ X_{t,T}
\}$ is stationary on the interval $[0,u_0]$.

To simplify our estimation method under conditions (i) and (ii), we
exploit the following fact: let $H_{(0)}$ and $\mathbb{H}_{(0)}$ be the
processes $H$ and $\mathbb{H}$ in the situation that $\{ X_{t,T} \}$ is
not only stationary up to $u_0$ but on the whole rescaled time interval
$[0,1]$. The covariance structure of $H_{(0)}$ is
$\operatorname{Cov}(H_{(0)}(u,v,f), H_{(0)}(u^\prime,v^\prime
,\break f^\prime)) =
\sigma_{(0)}^2(f,f^\prime)   C(u,u^\prime,v,v^\prime)$,
where\vspace*{1pt} $C(u,u^\prime,v,v^\prime)$ is defined in \eqref{known-cov} and
$\sigma_{(0)}^2(f,f^\prime) = \sum_{l=-\infty}^{\infty}
c_l(0,f,f^{\prime})$. Importantly, $H_{(0)}$ and $H$ have the
same covariance structure as long as $u,v,u^\prime,v^\prime\le u_0$.
From this, it is straightforward to see that the results of Section~\ref{subsec-asym-threshold} still hold true when the quantiles
$q_{\alpha}(u)$ of $\mathbb{H}(u)$ are replaced by those of $\mathbb{H}
_{(0)}(u)$. We may thus work with the quantiles of $\mathbb
{H}_{(0)}(u)$ in
Step 1(a) of the implementation.

Similarly as in Setting I, we now replace the statistic $\hat{\mathscr{D}}_T$
by a scaled version. Specifically, we scale it such as to get rid of
the unknown factor $\sigma_{(0)}^2(f,f^\prime)$ in the covariance
structure of $H_{(0)}$. To do so, let $\Sigma$ be the $K \times K$
matrix with the entries $\Sigma_{k,k^\prime} = \sigma
_{(0)}^2(f_k,f_{k^\prime})$. Supposing that $\Sigma$ is invertible, we
can write $\Sigma= A A^{\intercal}$ with a full-rank square matrix
$A$. The elements of $\Sigma$ can be estimated by $\hat{\Sigma
}_{k,k^\prime} = \hat{\sigma}_{(0)}^2(f_k,f_{k^\prime}) = \sum_{l=-\infty}^{\infty} W (\frac{l}{b(T)}) \hat
{c}_l(0,f_k,f_{k^\prime
})$, where $\hat{c}_l(0,f_k,f_{k^{\prime}}) = \sum_{t=l+1}^T
K_h(\frac
{t}{T}) \hat{Z}_{t,T}(f_k) \hat{Z}_{t-l,T}(f_{k^\prime}) / \sum_{t=1}^T
K_h(\frac{t}{T})$. In addition, we implicitly define $\hat{A}$ by
$\hat
{\Sigma} = \hat{A} \hat{A}^{\intercal}$. With this notation at hand,
we replace $\hat{D}_T$  by the scaled version\vspace*{-1pt}
\[
\bigl(\hat{D}_T^{\mathrm{sc}}(u,v,1),\ldots,
\hat{D}_T^{\mathrm
{sc}}(u,v,K)\bigr)^{\intercal}
= \hat{A}^{-1} \bigl(\hat{D }_T(u,v,f_1),
\ldots,\hat{D}_T(u,v,f_K) \bigr)^{\intercal},
\]
and define $\hat{\mathscr{D}}_T^{\mathrm{sc}}(u) = \max_{1 \le k \le
K} \sup_{v
\in[0,u]} |\hat{D}_T^{\mathrm{sc}}(u,v,k)|$.

When setting up our method in terms of $\hat{\mathscr{D}}_T^{\mathrm{sc}}$
rather than $\hat{\mathscr{D}}_T$, the quantiles in Step 1 of our
implementation need not be estimated but are known: by our
normalization, the limit process $H_{(0)}$ gets replaced by a scaled version $H
_{(0)}^{\mathrm{sc}}$, which has the covariance structure
\[
\operatorname{Cov}\bigl(H_{(0)}^{\mathrm{sc}}(u,v,k),H_{(0)}^{\mathrm
{sc}}
\bigl(u^\prime ,v^\prime,k^\prime\bigr)\bigr) =
\cases{ \ds C\bigl(u,u^\prime,v,v^\prime\bigr), & \quad$\mbox{if } k = k^\prime$,\vspace*{3pt}
\cr
0, & \quad $\mbox{if } k \ne
k^\prime$.}
\]
This covariance structure does not depend on any unknown quantities,\break
implying that the quantiles $q_{(0),\alpha}^{\mathrm{sc}}(u)$ of
$\mathbb{H}
_{(0)}^{\mathrm{sc}}(u)=\max_{1<k<K}\times\break \sup_{0\leq w\leq v\leq u}|H^{\mathrm{sc}}_{(0)}(v,w,k)|$   are known  (up to simulation error). Hence, when
constructing our estimator in terms of $\hat{\mathscr{D}}_T^{\mathrm
{sc}}$, we
can work with the known quantiles $q_{(0),\alpha}^{\mathrm{sc}}(u)$ in
Step 1 of the implementation.

\section{Finite sample properties}\label{sec-sim}

In this section, we examine the small sample performance of our
estimation procedure in a Monte Carlo experiment and compare it with
other estimation methods. In addition, we illustrate the methodology by
two real-data examples.

\subsection{Simulations}\label{subsec-sim}

We investigate two simulation setups: a time-varying mean model and a
volatility model together with a multivariate extension of it. Due to
space constraints, the results on the volatility models are presented
in the supplement [\citet{vogtdett2014}]. Here, we examine the setting
%
\begin{equation}
\label{mod1} X_{t,T} = \mu \biggl( \frac{t}{T} \biggr) +
\varepsilon_t
\end{equation}
with two different mean functions $\mu_1$ and $\mu_2$. The residuals
$\varepsilon_t$ are assumed to follow the AR(1) process $\varepsilon_t
= 0.25 \varepsilon_{t-1} + \eta_t$, where the innovations $\eta_t$ are
i.i.d. normal with zero mean and standard deviation $0.5$. The mean
functions are given by
%
\begin{eqnarray}
\mu_1(u) & = & 1(u > 0.5), \label{mu1a}
\\
\label{mu2a}
\mu_2(u) & =&  \bigl\{10(u-0.5)\bigr\} \cdot1(0.5 < u < 0.6) + 1(u >
0.6).
\end{eqnarray}
Both functions are equal to zero on the interval $[0,0.5]$ and then
start to vary over time. Hence $u_0 = 0.5$ in both cases. The function
$\mu_1$ is a step function which allows us to investigate how our
method works in the presence of abrupt changes. The function $\mu_2$ in
contrast varies smoothly over time. In particular, it starts to
linearly deviate from zero at the point $u_0 = 0.5$ until it reaches a
value of one and then remains constant.

\begin{figure}[b]

\includegraphics{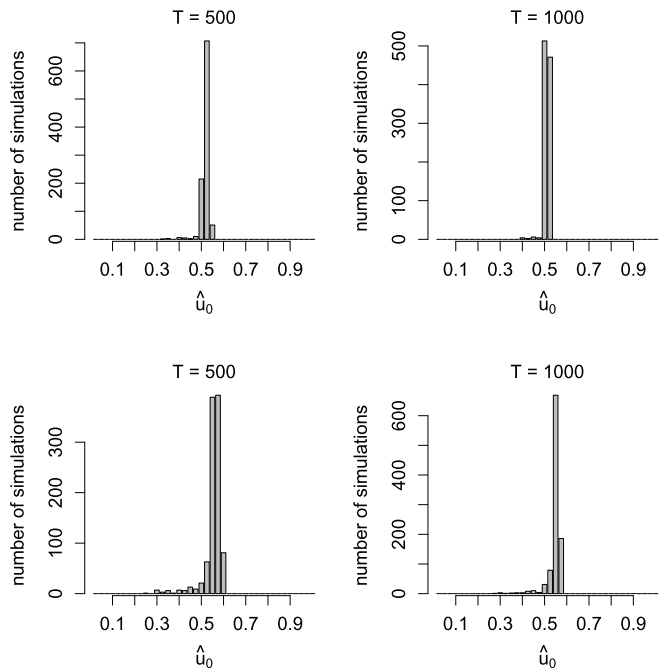}

\caption{Simulation results produced by our method in model \protect\eqref{mod1}. Upper panel: results for the mean function $\mu_1$ defined in
\protect\eqref{mu1a}. Lower panel: results for the mean function $\mu_2$
defined in \protect\eqref{mu2a}.}\label{fig-mu}
\end{figure}

To estimate the point $u_0$, we use the implementation strategy from
Setting I of Section~\ref{sec-impl} and denote the resulting estimator
by $\hat{u}_0$. We set the parameter $\alpha$ to equal $0.1$ in all our
simulations, meaning that the probability of underestimating $u_0$ is
approximately $10\%$. Moreover, as described in Setting I of Section~\ref{sec-impl}, we normalize the statistic $\hat{\mathscr{D}}_T(u)$
by an
estimate of the long-run error variance $\sigma^2 = \sum_{l=-\infty}^{\infty} \mathbb{E}[ \varepsilon_0 \varepsilon_l]$.
To do so, we
first approximate the residuals $\varepsilon_t$ by $\hat{\varepsilon}_t
= X_{t,T} - \hat{\mu}_h(\frac{t}{T})$, where $\hat{\mu}_h$ is a
Nadaraya--Watson estimator of $\mu$, and then apply a HAC estimator
with a Bartlett kernel to the estimated residuals. To construct $\hat
{\mu}_h$, we use an Epanechnikov kernel and set $h = 0.2$. Moreover, we
choose the bandwidth of the HAC estimator to equal $10$; that is, we
take into account the first ten autocovariances. As a robustness check,
we have repeated the simulations for different bandwidths. As this
yields very similar results, we have not reported them here.

For each model setting, we produce $N = 1000$ samples of length $T \in
\{ 500, 1000 \}$ and apply our procedure to estimate $u_0$. We thus
obtain $N=1000$ estimates of $u_0$ for each model specification. The
results are presented by histograms that show the empirical
distribution of the estimates for each specification. In particular,
the bars in the plots give the number of simulations (out of a total of
$1000$) in which a certain value $\hat{u}_0$ is obtained.

The simulation results for the design with $\mu_1$ are presented in the
upper part of Figure~\ref{fig-mu}, the left-hand panel corresponding to
a sample size of $T=500$ and the right-hand one to $T=1000$. Since $\mu
_1$ has a jump at $u_0 = 0.5$, it deviates from zero very quickly. Our
procedure is thus able to localize the point $u_0$ quite precisely.
This becomes visible in the histograms which show that the estimates
are not very dispersed but cluster tightly around $u_0=0.5$. The plots
also make visible a slight upward bias which becomes less pronounced
when moving to the larger sample size $T=1000$. The results for the
function $\mu_2$ are depicted in the lower part of Figure~\ref{fig-mu}.
The plots show that the upward bias is more pronounced than in the
setting with $\mu_1$. This reflects the fact that it is more difficult
to localize a gradual change than a jump.

In both designs, there is a certain fraction of estimates which takes
values below $u_0$. Theoretically, this fraction should be around $10\%
$, since we have set the probability $\alpha$ of underestimating $u_0$
to equal $0.1$. In our simulations, however, the fraction obviously
lies below the $10\%$ target as can be seen from the plots. This is a
small sample effect which can be explained as follows: our preliminary
estimate $\hat{u}_0^{\circ}$ is quite conservative, tending to strongly
overestimate $u_0$. Since $q_{\alpha}(u) \ge q_{\alpha}(u_0)$ for $u >
u_0$, this implies that the estimate $\hat{\tau}_{\alpha} =
q_{\alpha
}(\hat{u}_0^{\circ})$ will often overshoot the value of the critical
threshold $\tau_{\alpha} = q_{\alpha}(u_0)$, which underlies the second
step estimator $\hat{u}_0$. As a result, the empirical probability of
underestimating $u_0$ tends to lie below the target $\alpha$ in small samples.

\subsection{Comparison with other methods}

In this subsection, we compare our estimation approach with the methods
of \citet{Mallik2011} and \citet{huskova1999} which are specifically
designed to detect gradual changes in the location model~\eqref{mod1}.
As before, we assume that the mean function $\mu$ is constant on the
time interval $[0,u_0]$, that is, $\mu(u) = \overline{\mu}$ for $u
\le
u_0$, and then starts to vary over time. The method of \citet
{Mallik2011} allows one to estimate the time point $u_0$ when $\mu$ is
a smooth nonparametric function that is restricted to take values
larger than $\overline{\mu}$ at time points $u > u_0$, that is, $\mu(u)
> \overline{\mu}$ for $u > u_0$. The procedure of \citet{huskova1999} in
contrast is based on the parametric assumption that $\mu(u) =
\overline
{\mu} + \delta\cdot(u - u_0)^{\beta} \cdot1(u > u_0)$ for some slope
parameter $\delta> 0$ and a known constant $\beta\in[0,1]$. In what
follows, we set $\beta= 1$, thus considering Hu\u{s}kov\'{a}'s method
for the class of broken lines with a kink at $u_0$.

To compare our method with these two approaches, we set $u_0 = 0.5$ and
consider two different specifications of the mean function $\mu$,
%
\begin{eqnarray} \label{mu1}
\mu_4(u) & =& 2 (u - 0.5) \cdot1(u > 0.5),
\\
\label{mu2}
\mu_5(u) & =& \bigl\{10(u-0.5)\bigr\} \cdot1( 0.5 < u < 0.6) + 1 (u
\ge0.6).
\end{eqnarray}
Moreover, we let $\varepsilon_t$ be i.i.d. residuals that are normally
distributed with mean zero and standard deviation $0.2$. Note that $\mu
_4$ belongs to the parametric family of broken lines for which Hu\u
{s}kov\'{a}'s method with $\beta= 1$ is designed. The function $\mu
_5$, in contrast, is not an element of this parametric family.

Our estimator is implemented in the same way as in Section~\ref{subsec-sim}. As the error terms are i.i.d., the error variance
simplifies to $\sigma^2 = \mathbb{E}[\varepsilon_t^2]$ and can be
estimated as
follows: since $\mu$ is smooth, $\mu(\frac{t}{T}) - \mu(\frac{t-1}{T})
= O(T^{-1})$. This implies that $X_{t,T} - X_{t-1,T} = \varepsilon_t -
\varepsilon_{t-1} + O(T^{-1})$, which in turn yields that $\mathbb{E}(X_{t,T}
- X_{t-1,T})^2 = \mathbb{E}(\varepsilon_t - \varepsilon_{t-1})^2 +
O(T^{-2}) =
2 \sigma^2 + O(T^{-2})$. Hence, we may simply estimate the error
variance by $\hat{\sigma}^2 = T^{-1} \sum_{t=2}^T (X_{t,T} -
X_{t-1,T})^2/2$. This estimate is also used in the implementation of
the method by \citet{Mallik2011}. Hu\u{s}kov\'{a}'s estimator is
constructed as described in equation (1.4) of \citet{huskova1999}. To
implement the estimator of \citet{Mallik2011}, we proceed as follows:
since the method is based on a Nadaraya--Watson smoother of $\mu$, we
first select the bandwidth $h$ of this estimator. As shown in \citet
{Mallik2013}, the rate-optimal bandwidth has the form $h = c
T^{-1/(2k+1)}$, where $c$ is a constant and $\mu$ is assumed to have a
cusp of order $k$ at the point $u_0$. This means that the first $(k-1)$
right derivatives of $\mu$ at $u_0$ are zero, and the $k$th right
derivative is nonzero. For both functions, $\mu_4$ and $\mu_5$, $k$ is
equal to $1$, implying that the optimal bandwidth is of the form $h = c
T^{-1/3}$. Of course, since the order $k$ is unknown in practice, this
is not a feasible choice of bandwidth. Moreover, even if $k$ were
known, it is not clear how to pick the constant $c$. We here ignore
these problems and pretend that $k$ is known. Having repeated the
simulations for different choices of the constant $c$, we present the
results for the choice $c = 0.1$ which yields the best performance. For
simplicity, we also assume that the baseline value $\overline{\mu}$ is
known, so we do not have to replace it by an estimate.

\begin{figure}

\includegraphics{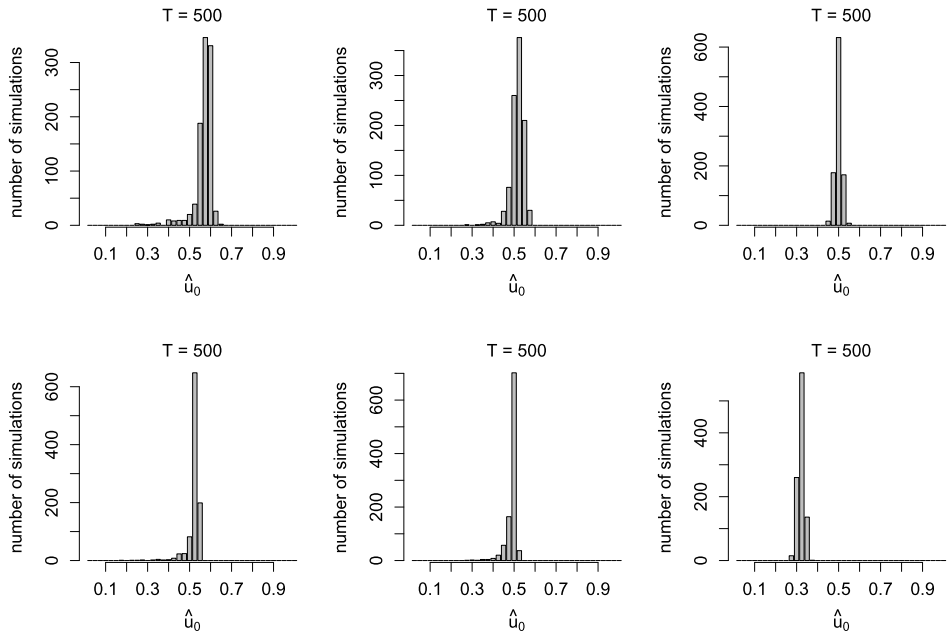}

\caption{Estimation results for model \protect\eqref{mod1} with $\mu_4$ (upper
panel) and $\mu_5$ (lower panel). The \mbox{left-hand} plots correspond to our
method, the middle ones to the approach of \protect\citet{Mallik2011} and the
right-hand ones to the procedure in \protect\citet{huskova1999}.}\label{fig-compare-1}
\end{figure}

The results for the regression function $\mu_4$ are presented in the
upper part of Figure~\ref{fig-compare-1}. As can be seen, Hu\u{s}kov\'
{a}'s method outperforms both ours and the $p$-value based approach of
\citet{Mallik2011}. This is not surprising since it is tailored to a
specific parametric class of mean functions to which $\mu_4$ belongs.
Even though less precise than Hu\u{s}kov\'{a}'s estimator, both our
method and the $p$-value based method perform well, ours tending to be
a bit more upward biased and thus slightly more conservative. The
results for the regression function $\mu_5$ are presented in the lower
part of Figure~\ref{fig-compare-1}. As before, both our method and that
of Mallik et al. perform quite well. The parametric method of \citet
{huskova1999}, in contrast, completely fails to provide reasonable
estimates of $u_0$. The reason for this is simply that $\mu_5$ does not
satisfy the parametric assumptions of this approach.

\begin{figure}

\includegraphics{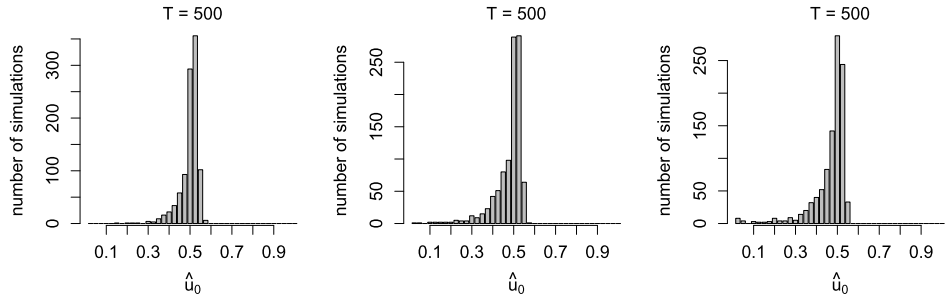}

\caption{Results for the method of \protect\citet{Mallik2011} in model \protect\eqref
{mod1} with $\mu_4$ and the bandwidth $h = c T^{-1/3}$, where $c = 0.2$
(left), $c = 0.3$ (middle) and $c = 0.4$ (right).}\label{fig-compare-2}
\end{figure}

To implement the method of \citet{Mallik2011}, we have used an optimally
tuned bandwidth which presupposes knowledge of the degree of smoothness
$k$, and we have treated the mean value $\overline{\mu}$ as known.
Nevertheless, this approach only provides slightly better results than
ours. In practice, $\overline{\mu}$ must of course be estimated, and
the optimal choice of bandwidth is not available. Moreover, the
performance of the method varies quite considerably with the bandwidth.
This is illustrated in Figure~\ref{fig-compare-2} which shows the
estimation results when picking the bandwidth to be rate optimal with
the constants $c = 0.2, 0.3, 0.4$ instead of $c=0.1$. As can be seen,
the results get much worse when slightly changing the bandwidth
parameter $c$, a~large fraction of the estimates tending to strongly
underestimate $u_0$. (Note that for all these values of $c$, the
bandwidth is fairly small, resulting in an undersmoothed estimate of
the mean function $\mu$. Specifically, for a sample size of $T=500$,
the choice $c = 0.1$ corresponds to a bandwidth window $Th$ of
approximately $6$ data points and $c = 0.4$ to a window of $25$ points.
Indeed, the method appears only to work in a reasonable way when
strongly undersmoothing, which is already indicated by the fact that
the optimal bandwidth is of the rate $T^{-1/3}$.)

The above discussion points to an important advantage of our method:
the tuning parameter $\tau_\alpha$ on which it depends is much more
harmless than a bandwidth parameter. As $\alpha$ can be interpreted in
terms of the probability of underestimating $u_0$, it is clear how to
choose $\tau_\alpha$ in a reasonable way in practice. Hence we do not
run the risk of producing poor estimates by picking the tuning
parameter in an inappropriate way. This makes our procedure
particularly attractive to apply in practice. We finally point out that
the new method is not specifically designed for detecting a change in
the nonparametric location model \eqref{mod1} but can be easily adapted
to other change point problems. This is illustrated in the supplement
[\citet{vogtdett2014}], where we show results for a nonparametric
volatility model.

\subsection{Applications}

We now apply our method to data from the 1997 Asian financial crisis.
Economists are interested in better understanding when the crisis hit
the different East Asian countries and when (and whether) the economic
shock spilt over from a particular country to another; see, for example,
\citet{Manner2014}. In what follows, we focus attention on the first
issue. To analyze it, we consider the daily stock index returns of
three countries that were badly hit by the crisis, namely Thailand,
Indonesia and Hong Kong. The three return series span the period from 7
January 1996 to 30 June 1998. Let $\{ r_{t,T}^{(i)} \}$ denote the
daily returns of the stock index of country $i$, and let $\lambda
_{t,T}^{(i)} = \mathbb{E}(r_{t,T}^{(i)})^2$ be the variance level at
the time
point $t$. We aim to estimate the time point when the variance level
starts to increase. To do so, we implement our method as described in
Setting~II of Section~\ref{sec-impl}. Specifically, we set $\alpha=
0.1$ and choose $h = 0.1$, noting that the results are very robust to
different choices of $h$. In addition, we set the bandwidth $b$ to
equal zero, assuming that the return data are independent over time.
\begin{table}[b]
\tablewidth=180pt
\caption{Estimated dates when the volatility levels of the stock
indices of Thailand, Indonesia and Hong Kong started to
increase}\label{table1}
\begin{tabular*}{180pt}{@{\extracolsep{\fill}}lc@{}}
\hline
Thailand & 02.09.1996 \\
Indonesia & 05.08.1997 \\
Hong Kong  & 02.09.1997 \\
\hline
\end{tabular*}
\end{table}

Table~\ref{table1} summarizes the results. As can be seen, Thailand was
first affected by the crisis. This is in accordance with the usual
opinion that the crisis broke out in Thailand and then spread over East
Asia. Our estimate suggests that the variance level starts to slowly
trend upwards already in September 1996, that is, already a few months
before the main events of the crisis. The variance levels of the other
two indices start to increase in August and September 1997,
respectively. The estimated dates can be linked to some important
events of the crisis. For example, the variance of the Indonesian stock
index started to go up in August 1997 when the Indonesian rupiah
strongly dropped and the IMF intervened.
In addition to the analysis of the individual variance levels, we
consider the data as a three-dimensional time series and estimate the
time point where the covariance matrix starts to change. Since the
covariance matrix is $3 \times3$ and symmetric, it has six different
elements, implying that the corresponding family of functions $\mathcal
{F}$ has six elements as well. Our estimate is 02.09.1996, which
coincides with the estimated date for the variance level of the Thai
index. Hence our analysis suggests that the covariance structure of the
three-dimensional system starts to change when the variance of the Thai
stock index begins to increase.

We next turn to the daily return data of the S\&P 500 index which are
depicted in the right-hand panel of Figure~\ref{fig1}. In recent years,
a variety of models for financial returns have been proposed that
produce the typical volatility patterns of return data by simple
nonstationarities. This contrasts with classical stationary GARCH
models where the patterns are generated by ARMA-type equations. A
simple locally stationary model for financial returns is given by the equation
%
\begin{equation}
\label{return-model} r_{t,T} = \sigma \biggl(\frac{t}{T} \biggr)
\varepsilon_t,
\end{equation}
where $r_{t,T}$ denotes the daily return, $\sigma$ is a time-varying
volatility function and $\varepsilon_t$ are i.i.d. residuals with zero
mean and unit variance. Model \eqref{return-model} has been studied in
a variety of papers; see \citet{Drees2003} and \citet{Fryzlewicz2006},
among others. In many situations, it is realistic to assume that the
volatility level is more or less constant within some time span
$[u_0,1]$, where $u=1$ is the present time point,  and remains roughly
constant in the near future $(1,1+\delta]$. In this case, $\sigma(u)
\approx\sigma(1)$ at future time points $u \in(1,1+\delta]$, which
suggests that one should use the present volatility level $\sigma(1)$
as a forecast for the near future; see \citet{Fryzlewicz2006}, among
others. To obtain a good volatility forecast, we thus have to construct
a good estimator of $\sigma(1)$. If we knew the time point $u_0$, we
could come up with a very simple and precise estimator. In particular,
we could estimate $\sigma^2(1)$ by the sample variance of the
observations contained in the time interval $[u_0,1]$. In practice,
however, the time span $[u_0,1]$ is not observed but has to be estimated.

We now estimate the time span $[u_0,1]$ where the volatility level of
the S\&P 500 returns from Figure~\ref{fig1} is more or less constant.
To do so, we have to reformulate our estimation method, since it is
designed to apply to time spans of the form $[0,u_0]$ rather than
$[u_0,1]$. Since this is trivial to achieve and simply a matter of
notation, we neglect the details. As time-variation in the volatility
is equivalent to time-variation in the variance $\operatorname
{Var}(r_{t,T}) = \mathbb{E}
[r_{t,T}^2]$, we set up our procedure to detect changes in the variance
and implement it as described in Setting II of Section~\ref{sec-impl}.
As before, we let $\alpha= 0.1$, $h = 0.1$ and $b = 0$, again noting
that the results are robust to different choices of the bandwidth $h$.
Our estimate $\hat{u}_0$ of the time point $u_0$ is depicted as the
vertical dashed line in the right-hand panel of Figure~\ref{fig1}.

\section{Extensions}\label{sec-ext}

$\!$Even though our method is designed to detect gradual changes, it is
also able to detect abrupt breaks. Indeed, it does not only work for
locally stationary processes, that is, for processes whose stochastic
properties vary smoothly over time. It is valid for a much wider class
of nonstationary processes whose stochastic properties may change
gradually as well as abruptly. Generally speaking, we can deal with any
type of nonstationary behavior that allows us to derive the weak
convergence result on $\hat{H}_T$ in Theorem~\ref{theo-measure}.
Inspecting the supplement [\citet{vogtdett2014}], the proof of Theorem~\ref
{theo-measure} can be seen to split up into two parts:
\begin{longlist}[(2)]
\item[(1)] In Proposition A.2, we show that the process $\hat{H}_T$ is
asymptotically stochastically equicontinuous. Importantly, we do not
need the process $\{ X_{t,T} \}$ to be locally stationary to prove this
result. We only require appropriate mixing and moment conditions on $\{
X_{t,T} \}$, as imposed in (C2)--(C4), and can drop the local
stationarity condition (C1).
\item[(2)] In Proposition A.1, we show that the finite dimensional marginals
of the process $\hat{H}_T$ are asymptotically normal. This is the only
part of the proof where the local stationarity condition is needed. The
idea of the proof of Proposition~A.1 is to successively replace the nonstationary
variables $X_{t,T}$ by their stationary approximations $X_t(u)$. This
allows us to show that the covariance matrices of the marginals
converge to the limit expression \eqref{cov-H}. Hence the local
stationarity property is only required to ensure that the covariances
of the marginals are asymptotically well behaved.
\end{longlist}
Consequently, we can allow for any nonstationary behavior of $\{
X_{t,T} \}$ which ensures that the marginals of $\hat{H}_T$ have
covariances with a well-behaved limit. In particular, we may assume the
process $\{ X_{t,T}\dvtx  1 \le t \le T \}_{T=1}^{\infty}$ to be piecewise
locally stationary in the following sense: there exists a finite number
of time points $0 = v_0 < v_1 < \cdots< v_N = 1$ such that $\{
X_{t,T}\dvtx  v_{j-1} T \le t < v_j T \}_{T=1}^{\infty}$ is a locally
stationary process for each $j = 1,\ldots,N$. This definition allows
for a finite number of abrupt changes at the points $v_j$. For example,
we may consider the mean setting $X_{t,T} = \mu(\frac{t}{T}) +
\varepsilon_t$, where the function $\mu$ is smooth within the intervals
$[v_{j-1},v_j)$ but has jumps at the points $v_j$. Our method is able
to detect the time point $u_0$ where the mean function $\mu$ starts to
vary over time, no matter whether the change point $u_0$ is gradual
[i.e., $u_0 \in(v_{j-1},v_j)$ for some $j$] or abrupt (i.e., $u_0 =
v_j$ for some $j$).

\section*{Acknowledgments}
We would
like to thank Rainer Dahlhaus, Wolfgang Polonik and Stanislav Volgushev
for helpful discussions and comments on an earlier version of this
manuscript. We are also grateful to Alina Dette and Martina Stein, who
typed parts of this paper with considerable technical expertise. The
constructive comments of an Associate Editor and two referees on an
earlier version of this paper led to a substantial improvement of the
manuscript. Parts of this paper were written while the authors were
visiting the Isaac Newton Institute, Cambridge, UK, in 2014
(``Inference for change-point and related processes''), and the authors
would like to thank the institute for its hospitality.

\begin{supplement}[id=suppA]
\stitle{Supplement to ``Detecting gradual changes in locally stationary
processes''}
\slink[doi]{10.1214/14-AOS1297SUPP} 
\sdatatype{.pdf}
\sfilename{aos1297\_supp.pdf}
\sdescription{In the supplement, we examine the finite sample
performance of our method by further simulations. In addition, we
provide the proofs that are omitted in the paper.}
\end{supplement}



\printaddresses
\end{document}